\documentstyle[12pt,a4,epsfig,here]{article}

\def\nextline{\hfill\break}

\def\mycomm#1{\nextline\strut\kern-3em{\tt ====> #1}\nextline}

\def\gray{\special{ps: 0.4 setgray}}
\def\black{\special{ps: 0.0 setgray}}

\newcommand{\draft}{
\newcount\timecount
\newcount\hours \newcount\minutes  \newcount\temp \newcount\pmhours
 
\hours = \time
\divide\hours by 60
\temp = \hours
\multiply\temp by 60
\minutes = \time
\advance\minutes by -\temp
\def\hour{\the\hours}
\def\minute{\ifnum\minutes<10 0\the\minutes
            \else\the\minutes\fi}
\def\clock{
\ifnum\hours=0 12:\minute\ AM
\else\ifnum\hours<12 \hour:\minute\ AM
      \else\ifnum\hours=12 12:\minute\ PM
            \else\ifnum\hours>12
                 \pmhours=\hours
                 \advance\pmhours by -12
                 \the\pmhours:\minute\ PM
                 \fi
            \fi
      \fi
\fi
}
\def\fullclock{\hour:\minute}
\begin{centering}
\gray
\special{ps: -90 rotate}
\special{ps: -4600 -5100 translate}
\font\Hugett  =cmtt12 scaled\magstep4
{\Hugett Draft: \today, \clock}
\black
\special{ps: 90 rotate}
\special{ps: 5100 -4600 translate}
\end{centering}
\vskip -1.7cm
$\phantom{a}$
} 

\newcommand{\bmath}{\begin{displaymath}}
\newcommand{\emath}{\end{displaymath}}
 
\def\beq{\begin{equation}}
\def\eeq{\end{equation}}
 
\newcommand{\bea}{\begin{eqnarray}}
\newcommand{\eea}{\end{eqnarray}}
 
\def\eqref#1{(\ref{#1})}

\newcounter{saveeqn}

\catcode`\@=11 

\def\eqarraylabel#1{\@bsphack \if@filesw 
{\let \thepage \relax \def \protect {\noexpand \noexpand \noexpand }%
\edef \@tempa {\write \@auxout {\string \newlabel 
{#1}{{\mbox{\arabic{saveeqn}}}{\thepage }}}}\expandafter }\@tempa 
\if@nobreak \ifvmode \nobreak \fi \fi \fi \@esphack}

\catcode`\@=12 
 
\newcommand{\la}{$\Lambda$ }
\newcommand{\lab}{$\bar{\Lambda}$ }

\def\tsize{\Large} 
\def\asize{\normalsize} 
\title {
\begin{flushright}
\normalsize TAUP - 2621-2000\\
\normalsize WIS2000/3/Feb.-DPP
\end{flushright}
\vspace{2.0cm}
\tsize
The contribution of $\Sigma^* \rightarrow \Lambda\pi$ to measured \la
polarization\thanks{Supported
in part by grants from US-Israel Bi-National Science Foundation
and from the Israeli Science Foundation.}
}
\author{\asize
D. Ashery$^{1}$ \\
\asize and\\
\asize $\phantom{a}$ H. J. Lipkin$\,^{1,2}$
\vspace{0.5cm}
\\
\asize \sl $^1$\,School of Physics and Astronomy\\
\asize \sl The Raymond and Beverly Sackler Faculty of Exact Sciences\\
\asize \sl Tel Aviv University, 69978 Tel Aviv, Israel\\
\asize \sl and\\
\asize \sl $^2$\,Department of Particle Physics\\
\asize \sl The Weizmann Institute of Science, 76100 Rehovot, Israel\\
}
\date { }
\begin {document}
\maketitle
\begin{abstract}
\noindent
Calculations of the polarization of \la and \lab particles after fragmentation 
of a polarized quark produced in processes like 
$Z$-decay and deep inelastic polarized lepton scattering must 
include \la and \lab produced as decay products of $\Sigma^0$ and
$\Sigma^*$ as well as those produced directly. These decay contributions
are significant and not feasibly included in theoretical calculations
based on QCD without additional input from other experimental data.
Furthermore these contributions depend on the spin structure of the
$\Sigma^0$ or $\Sigma^*$ and are not directly related to the structure
function of the $\Lambda$ 
 \end{abstract}
\thispagestyle{empty} 
\draft
\newpage
The interpretation of studies of the nucleon spin structure functions
is that quarks in the
nucleon carry only $\sim$30\% of the nucleon spin and that the strange   
(and non-strange) sea is polarized opposite to the polarization of the
valence quarks. An attempt to shed light on this very problematic
conclusion was made through 
measurement of the polarization of \la and \lab produced near the Z pole
in $e^+e^-$ collisions \cite{aleph,opal} and in polarized lepton Deep
Inelastic Scattering on unpolarized targets \cite{e665,herm}.
Several theoretical works were published on this subject in which
predictions and calculations relevant to the interpretation of the
experimental results were presented
\cite{burjaf,jaf,def,thom,ekks,elliskk,koz,dazl}.
A difficulty present in most experiments is that they cannot distinguish
between \la and \lab
produced directly or as decay
products. The main contribution from decays is the $\Sigma^* \rightarrow
\Lambda\pi$. The purpose of the present note is to emphasize the
importance of taking into account the contribution from this process.\\

To illustrate this point, if 100\% polarized strange quarks hadronize
directly to a \la, the \la polarization will be also 100\% if the spin
structure of the \la is as expected by the na\"{\i}ve quark model. On the
other hand, if they hadronize to a $\Sigma^*$ the \la particles resulting
from its decay will be only 55\% polarized \cite{dazl}. If the spin
structure of the \la is as derived from SU(3) symmetry using the
measured spin structure of the proton, 100\% strange quarks will result in
73\% polarized \la if produced directly compared with the same 55\% if
coming from $\Sigma^*$ decay.\\

The polarization of the $\Lambda$ particles observed in any experiment can
be written
\beq
P (\Lambda) = 
{{N_{nd} \cdot P_{nd}(\Lambda) 
+ N_{\Sigma^*}\cdot BR(\Sigma^* \rightarrow \Lambda \pi)\cdot 
P_{\Sigma^*}(\Lambda) 
+  N_{\Sigma^o}\cdot P_{\Sigma^o}(\Lambda) 
}\over{N_{\Lambda} }}
\label{QQ101}
\eeq
where $N_\Lambda$, $N_{\Sigma^*}$ and 
$ N_{\Sigma^o}$ denote respectively the 
numbers of $\Lambda$'s, $\Sigma^*$'s and 
$ \Sigma^o$'s produced in the experiment, $BR(\Sigma^* \rightarrow \Lambda \pi)$
denotes the
branching ratio for the $\Sigma^* \rightarrow \Lambda \pi$ decay,
$ P_{\Sigma^*}(\Lambda)$  and 
$P_{\Sigma^o}(\Lambda)$ 
denote respectively the polarizations of the 
$\Lambda$'s produced via the $\Sigma^* \rightarrow \Lambda \pi$ decay and 
the 
$\Sigma^o \rightarrow \Lambda \gamma$ decay, and $N_{nd}$ and 
$P_{nd}(\Lambda)$ denote the number and polarization of $\Lambda$'s produced 
via all other 
ways; i.e. which  which do not go via the $\Sigma^*$ or $ \Sigma^o$, 
\beq
N_{nd} = N_\Lambda - N_{\Sigma^*}\cdot BR(\Sigma^*
\rightarrow \Lambda \pi) - N_{\Sigma^o}
\label{QQ107}
\eeq

The individual terms in the numerator of eq.(\ref{QQ101}) are all distinct and
measurable. Any calculation of the polarization of the final observed 
$\Lambda$ must consider all these contributions if they are not separated 
experimentally. 

One might argue that the $\Sigma^* \rightarrow \Lambda \pi$ decay is a strong
interaction described in terms of quarks and gluons in QCD and should be
included in the inclusive polarized fragmentation function. Clearly the
$\Sigma^*$ intermediate state must be already included in any fragmentation
function which takes into account $all$ strong interactions in the description
of a process in which a struck quark turns into a $\Lambda$ plus anything else.
This point of view is implied in the treatments \cite{burjaf,jaf} which
attempt
to use the $\Lambda$ polarization data to extract fragmentation functions. But
the expression for the $\Lambda$ polarization eq.(\ref{QQ101}) is rigorous.
Thus a theoretical formulation which gives a prediction for this polarization
must also include a prediction for the precise values for all parameters
appearing in eq.(\ref{QQ101}) 

We immediately find a crucial weak point in all attempts to obtain a
theoretical estimate for the value of the $\Lambda$ polarization. Any
theoretical attempt to obtain the value of the branching ratio $BR(\Sigma^*
\rightarrow \Lambda \pi)$ must take into account fine threshold effects like the
small SU(3) breaking produced by the $\Lambda-\Sigma$ mass difference which
vanishes in the SU(3) symmetry limit. Note that in the SU(3) symmetry
limit the predicted ratio of the branching ratios of the two $\Sigma^*$
decay modes is in strong disagreement with experiment:
\beq
\left({{ BR(\Sigma^* \rightarrow \Lambda \pi)}\over{BR(\Sigma^* \rightarrow 
\Sigma \pi)}}\right)_{theo} = 1/2 \not= 
\left({{ BR(\Sigma^* \rightarrow \Lambda \pi)}\over{BR(\Sigma^* \rightarrow 
\Sigma \pi)}}\right)_{exp} = 7.3 \pm 1.2
\label{QQ102}
\eeq
The disagreement is more than an order of magnitude.

There is also the problem of obtaining the values of $N_{nd}$, $N_{\Sigma^*}$ 
and $ N_{\Sigma^o}$. If one assumes a purely
statistical model in which all states of two nonstrange quarks and one
strange quark are equally probable the ratio 
\beq
N_{nd}/N_{\Sigma^*}/N_{\Sigma^o} = 1:6:1
\label{QQ112}
\eeq
where the factor 6 arises from the $(2J+1)$ spin factor and the three charge
states of the $\Sigma^*$  which all decay into $\Lambda -\pi$. The experimental
values are very different. There is also the problem that all three charge
states are equally produced by the fragmentation of a struck $s$ quark, while a
struck $u$ or $d$ quark can only produce the two charge states containing the
struck quark. All these factors complicate any attempt at this stage to predict
values of $N_{nd}$, $N_{\Sigma^*}$ and $ N_{\Sigma^o}$ from any purely
theoretical model without any external experimental input. 

Thus predictions for polarization of the $\Lambda$'s observed in any experiment
must include as input the known experimental branching ratio $BR(\Sigma^*
\rightarrow \Lambda \pi)$ as well as the values of $N_{nd}$, $N_{\Sigma^*}$ and
$ N_{\Sigma^o}$ obtained from other experiments or from Monte Carlo programs
which rely on a number of free parameters which are adjusted to fit vast
quantities of data. Note that the $\Sigma^o$ decays electromagnetically. Its
decay is never included in any strong interaction fragmentation function and
the $\Lambda's$ produced via its production and decay must be considered
separately in all fragmentation models. 

We now note that the  polarization of \la produced from the decay
of a $ \Sigma^* $ or $\Sigma^o $ is proportional to the polarization of
the decaying $ \Sigma^* $ or $\Sigma^o $ with coefficients depending only
on  angular momentum Clebsch-Gordan coefficients and completely
independent of the spin structure of the $\Lambda$\cite{dazl}:
\beq
P_{\Sigma^*}(\Lambda) = P_{\Sigma^*}\cdot C(\Sigma^*); ~ ~ ~ 
P_{\Sigma^o}(\Lambda) = P_{\Sigma^o}\cdot C(\Sigma^o); ~ ~ ~ 
P_{\Sigma^*\Sigma^o }(\Lambda)  = P_{\Sigma^*}\cdot C(\Sigma^* \rightarrow 
\Sigma)
\label{QQ103}
\eeq
where $P_{\Sigma^*}$ and $P_{\Sigma^o}$ denote the polarizations respectively
of the $\Sigma^*$ and $\Sigma^o$ before their decays, and $C(\Sigma^*)$,
$C(\Sigma^o)$ and $C(\Sigma^* \rightarrow \Sigma)$ denote the model-independent
functions of Clebsch-Gordan coefficients describing the ratio of the
polarization of the final $\Lambda$ to the polarization of the decaying baryon.
The explicit values of these functions are given in ref. \cite{dazl}, where it
is shown that the polarization of the final $\Lambda$ in all models for the
baryons and the dynamics of the decay process is given by the polarization of
the strange quark in the simple constituent quark model for the decaying
baryon. We immediately note that only the polarization of the
directly-produced $\Lambda$ can depend upon the spin-flavor structure of the
$\Lambda$. The other terms in eq.(\ref{QQ101})  depend upon the spin-flavor
structure of the $\Sigma^*$ and the $\Sigma^o$, but are independent of the
spin-flavor structure of the $\Lambda$. 

The expression eq.(\ref{QQ101}) for the polarization of all the $\Lambda$'s
produced in a given experiment is easily generalized to obtain the polarization
of $\Lambda$'s restricted to a given domain of various kinematic variables. It
is necessary to note that the momenta of $\Lambda$'s produced from a decay of a
$\Sigma^*$ or $\Sigma^o$ are different from those of the parent baryon. Thus to
obtain the polarization of $\Lambda$'s produced in a given kinematic range one
must integrate the expressions for $N_{\Sigma^*}$ and $ N_{\Sigma^o}$ over
momenta with the appropriate weighting factors and angular distributions
needed to produce the $\Lambda$'s in the correct kinematic range. All this
cannot be done reliably in present theoretical calculations and has to be
taken from Monte Carlo simulations that are tuned and tested and thus
reproduce well many related experimental observables.

The polarization of \la and \lab produced at the Z-pole \cite{aleph,opal}
was calculated
using two ingredients: the polarization of the $s$ and $\bar{s}$ quarks
produced at the pole and their hadronization into \la and \lab. The first
part was derived from weak interactions \cite{burjaf} and can be predicted
directly. The hadronization of the $s,\bar{s}$  directly to \la and \lab
or as decay products was calulated using Monte Carlo simulations. The
authors found that about 20\% of the \la polarization is contributed by
\la particles originating from $\Sigma^*$ decay. These calculations used
the na\"{\i}ve quark model wave functions for the baryons. This shows that
inclusion of this contribution was essential for obtaining agreement with
the data. In any attempt to go beyond the na\"{\i}ve quark model and
include the information obtained from DIS on the spin-flavor structure of
the proton, it is clearly necessary not only to have a model for the
spin-flavor structure of the $\Lambda$, but also of the $\Sigma^*$ and the
$\Sigma^o$ as well.\\

\begin{figure}[H]
\epsfig{figure=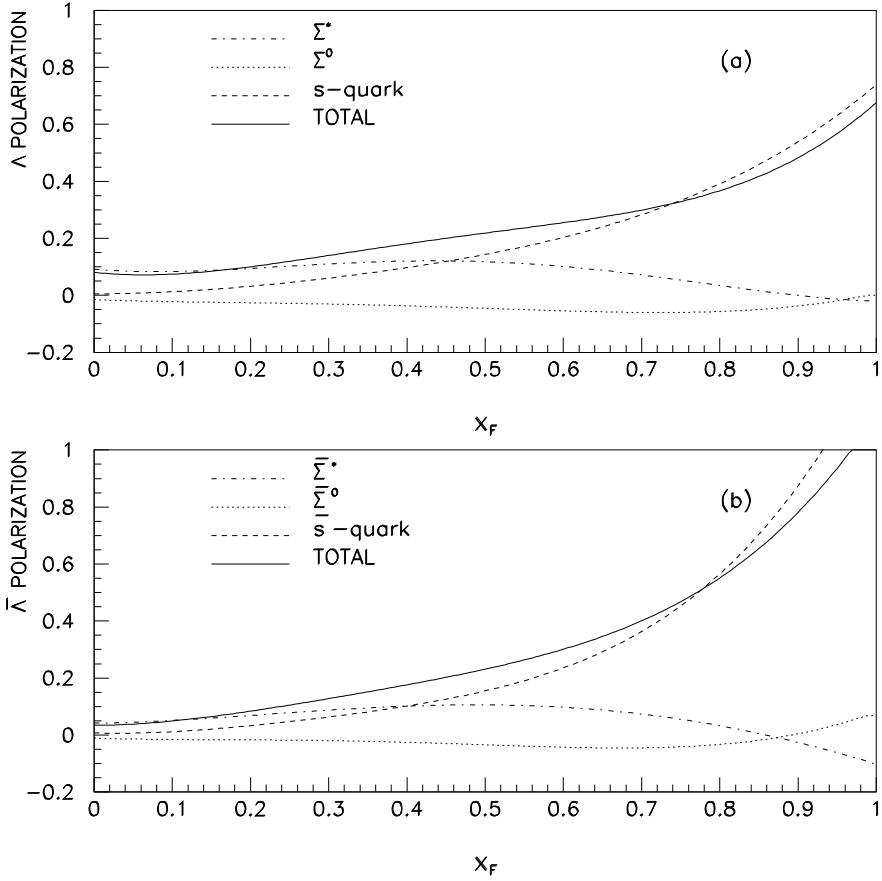,width=13cm}
\caption{Polarization of \la (a) and \lab (b) hyperons. Contributions from
direct production (dashed line), from decays of $\Sigma^0$ (dotted line),
of $\Sigma^*$ (dash-dotted line) and the total
polarization (solid line). All are calculated using the  na\"{\i}ve quark
model.}
\label{calc}
\end{figure}

In studies of \la
polarization in deep inelastic scattering of polarized muon
\cite{e665,dazl} the $x_F$ dependence of this contribution is studied
(see fig. \ref{calc}). It is found that up to $x_F \sim 0.5$ the
contribution from $\Sigma^*$ decay to the \la polarization is
dominant. It is only for larger $x_F$ values that polarization
from directly produced \la's becomes dominant. It should be noted that the
\la yield is dropping fast for large $x_F$ and consequently most of the
data is taken in the region where the \la polarization is dominated by
$\Sigma^*$ decay. The contribution from this process must therefore be
considered very carefully. This contribution was addressed only in some of
the theoretical calculations of this process \cite{elliskk,koz,dazl}.

\bibliographystyle{unsrt}

\end{document}